\documentclass{aa}

\usepackage{amsmath}
\usepackage{natbib,twoopt}
\bibpunct{(}{)}{;}{a}{}{,}             
\makeatletter
  \newcommandtwoopt{\citeads}[3][][]{\href{http://adsabs.harvard.edu/abs/#3}%
    {\def\hyper@linkstart##1##2{}%
     \let\hyper@linkend\@empty\citealp[#1][#2]{#3}}}
  \newcommandtwoopt{\citepads}[3][][]{\href{http://adsabs.harvard.edu/abs/#3}%
    {\def\hyper@linkstart##1##2{}%
     \let\hyper@linkend\@empty\citep[#1][#2]{#3}}}
  \newcommandtwoopt{\citetads}[3][][]{\href{http://adsabs.harvard.edu/abs/#3}%
    {\def\hyper@linkstart##1##2{}%
     \let\hyper@linkend\@empty\citet[#1][#2]{#3}}}
  \newcommandtwoopt{\citeyearads}[3][][]%
    {\href{http://adsabs.harvard.edu/abs/#3}
    {\def\hyper@linkstart##1##2{}%
     \let\hyper@linkend\@empty\citeyear[#1][#2]{#3}}}
\makeatother

\usepackage{gensymb}
\usepackage{graphicx}
\usepackage{txfonts}
\usepackage{gensymb}
\usepackage{multirow}
\usepackage{xspace}
\usepackage{amsmath}

\newcommand{\hh}{$\mathrm{H_2}$\xspace}

\begin{document}

   \title{{The magnetic properties of the protostellar core IRAS 15398-3359}}


   \author{E. Redaelli
          \inst{1}
          \and
         F. O. Alves\inst{1} \and
          F. P. Santos \inst{2} \and
        	 P. Caselli \inst{1} }
           
  \institute{Centre for Astrochemical Studies, Max-Planck-Institut f\"ur extraterrestrische
              Physik, Gie\ss enbachstra\ss e 1, D-85749 Garching bei M\"unchen (Germany) \\
              \email{eredaelli@mpe.mpg.de}
              \and
              Max-Planck-Institute for Astronomy, K\"onigstuhl 17, 69117 Heidelberg, Germany }

     \titlerunning{{Magnetic field in IRAS15398}}
     
   \authorrunning{Redaelli et al.}
   \date{Received ****; accepted *****}

 
  \abstract
   {Magnetic fields can affect significantly the star formation process. The theory of the magnetically-driven collapse in a uniform field predicts that initially the contraction happens along the field lines. When the gravitational pull grows strong enough, the magnetic field lines pinch inwards, giving rise to a characteristic hourglass shape.}
   {We investigate the magnetic field structure of a young Class 0 object, IRAS 15398-3359, embedded in the Lupus I cloud. Previous observations at large scales suggest that this source evolved in an highly magnetised environment. This object thus appears an ideal candidate to study the magnetically driven core collapse in the low-mass regime.}
   {We have performed polarisation observations of IRAS 15398-3359 at $214\,\mu$m using the SOFIA/HAWC+ instrument, thus tracing the linearly polarised thermal emission of cold dust.}
   { {Our data unveil a significant bend of the magnetic field lines due to the gravitational pull.} The magnetic field appears ordered and aligned with the large-scale B-field of the cloud and with the outflow direction. We estimate a magnetic field strength of {$B ={78}\,\mu$G}, expected to be accurate within a factor of two. The measured mass-to-flux parameter is {$\lambda = {0.95}$}, indicating that the core {is in a transcritical regime}.}
   {}


   \maketitle
%

\section{Introduction}
Magnetic fields  ($B$) are expected to play an important role in the star formation process, for instance providing a source of non-thermal pressure against the gravitational pull (see e.g. \citealt{McKee07}). High-density molecular clouds are threaded by magnetic fields aligned perpendicularly to the clouds' main axes \citep{PlanckXXXV}, which is the expected configuration when $B$-fields regulate the star formation process more effectively than turbulence \citep{Mouschovias76, Nakamura08, Li13}. At smaller scales ($\approx 0.1-1\,$pc), the collapse of spherical cores in uniform fields have been widely investigated \citep{Mouschovias91,Galli93,Allen03}. Theory predicts the formation of a flattened structure (\textit{pseudodisc}) because the collapse happens preferentially along the field lines. Given that the interstellar gas is often slightly ionised, matter is expected to be coupled with the magnetic field lines at envelope scales. Therefore, magnetic lines pinch inwards due to the gravitational pull, exhibiting a characteristic hourglass shape. In low-mass star forming regions this feature has been detected only in 30\% of Young Stellar Objects in polarisation (9 sources out of 32, \citealt{Hull19}), suggesting that this is not a universal picture. {Furthermore, out of these nine detections only two show a clear hourglass shape, namely IRAS 4A \citep{Girart06} and L1448 \citep{Kwon18}.}  
Polarisation observations in the far-infrared (FIR)/sub-millimeter wavelength are an effective way to investigate the magnetic properties of clouds and cores at intermediate/high visual extinctions ($A_\mathrm{V} \gtrsim 10 \,$mag). In fact, asymmetric dust grains illuminated by an external radiation field are expected to develop a magnetic moment and thus to align with their minor axes parallel to the magnetic field direction (radiative torque alignment, or RAT, \citealt{Lazarian07}). As a result, the background starlight at near-infrared/optical frequency is absorbed and linearly polarised in the $B$-field direction. Similarly, since the dust grains emit preferentially in the direction of their major axes, dust thermal emission is linearly polarised perpendicularly to the local $B$-field. Cold dust ($T \approx 10-30\,$K) emission happens typically at FIR/sub-mm wavelengths. \par
 IRAS 15398-3359 (hereafter IRAS15398) is a low-mass Class 0 protostar \citep{Andre00} located in the Lupus I molecular cloud, at a distance of $156\,$pc \citep{Dzib18}. {The protostellar mass, estimated from modelling of the envelope rotation, is subsolar ($M_* < 0.1\, M_\odot$, \citealt{Oya14, Yen17}).}  Lupus I is the least evolved cloud of the Lupus complex \citep{Rygl13}, and optical polarisation observations have shown that it is threaded by a very ordered magnetic field, perpendicular to its whole filamentary extension \citep{Franco15}.  IRAS15398 has a luminosity of $1.8\, L_\odot$ and an envelope mass of $1.2\, M_\odot$. It powers a bipolar outflow detected in CO lines both with single dish and interferometric observations \citep{Tachihara96, Bjerkeli16}.  \par
Using the SOFIA telescope, we have observed the polarised dust thermal emission arising from the core and the filamentary structure in which IRAS15398 is embedded. For the first time, we can compare the large scale field with the small scale field in a Class 0 source In this {paper}, we report our findings, which unveil {an ordered magnetic field, with hints of pinching of the field lines due to the gravitational pull.} Our results suggest that IRAS15398 evolved in an highly magnetised environment, and experienced a magnetically-driven collapse.

\section{Observations}
\begin{figure*}
\centering
\includegraphics[width = 0.7\textwidth]{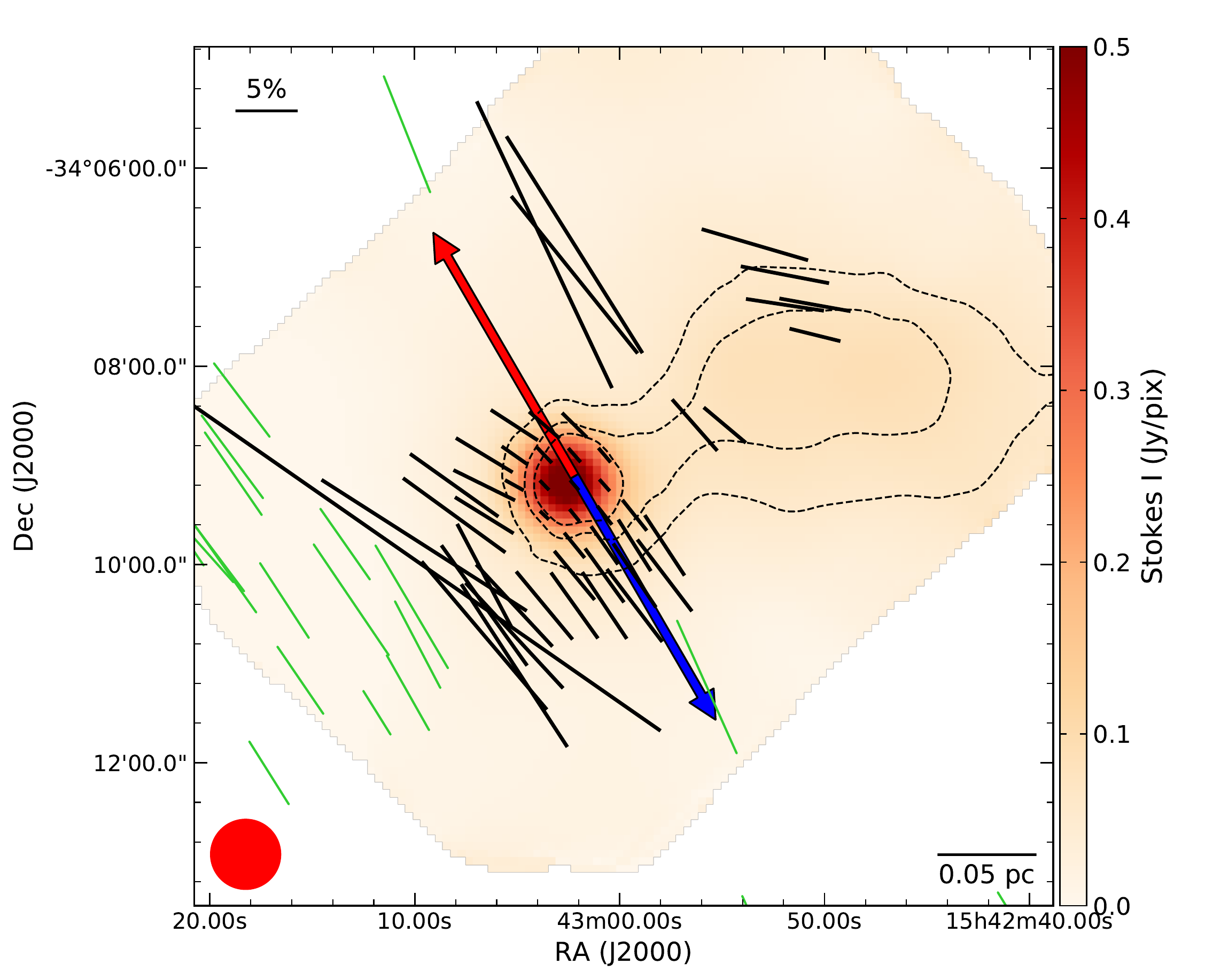}
\caption{Colorscale map of the Stokes I flux observed by SOFIA. The dashed contours represent $N\rm(H_2)$ column density as derived from {\em Herschel} data ($\rm levels = [1.0, \,1.5,\, 2.0]\times 10^{22} \,  cm^{-2}$). The black vectors show the polarization angles, tilted by 90\degree to trace the magnetic field direction, and with their length proportional to the polarization fraction (scalebar in the top-left corner). To show statistically independent data, we plot only two vectors per smoothed beam. The smoothed  beam size is shown in the bottom-left corner. In green, we show the polarisation vectors from optical observations, from \cite{Franco15}. The red/blue arrows indicate the direction of the outflow.\label{StokesIVec}}
\end{figure*}

\subsection{Polarimetric data}
The polarimetry observations were carried out with the SOFIA telescope, using the HAWC+ instrument \citep{Harper18} in summer 2018. We used the band E, with a nominal wavelength of $214\,\mu$m ($1.4\,$THz). At this frequency, the  SOFIA beam full-width-half-maximum is $FWHM\approx 19''$ (corresponding to $\approx 0.01\,$pc at the source's distance), and the field of view {is $5.0' \times 6.3'$}. The data were processed with the standard pipeline (HAWC DRP, version 1.3.0), and we used the flux-calibrated {Level 4 data products (already flux calibrated and mosaiced)}. The total integration time was $\approx \mathrm{1h15min}$, and the sensitivity achieved in the Stokes I is {$65\,$mJy/beam}. In order to increase the signal-to-noise ratio (SNR) of the data, and to have significant detections {over a large} fraction of the source, we smoothed the Stokes parameter I, Q, and U to $42''$ resolution ($\approx 0.03\,$pc), {before producing maps of polarization intensity and position angle (see Sec. \ref{Results})}. The final mean sensitivity is {$rms = 48\,$mJy/beam (Stokes I), $77\,$mJy/beam (Stokes Q), and $77\,$mJy/beam} (Stokes U). Since we found no indication of spatially correlated noise, the smoothing technique as a tool to increase SNR is justified. 

\subsection{$\mathrm{H_2}$ column density maps}
To obtain the gas column density $N\rm(H_2)$ and the visual extinction $A_\mathrm{V}$ maps of the source, we use the archive data of the Gould Belt Survey, performed with the {\em{Herschel}} space telescope \citep{Andre10}. They provide {directly} the \hh column density map of Lupus I, obtained with a spectral fitting of the dust emission at 5 wavelengths (70, 160, 250, 350, and $500\mu$m). This map has a resolution of $\approx 38''$, which allows a fair comparison with the SOFIA smoothed data. We derived the visual extinction map using the standard relation $A_\mathrm{V}/ \mathrm{mag}= 1.06\times10^{-21} \cdot N\mathrm{(H_2)}/\mathrm{cm^{-2}} $, \citep{Bohlin78}.

\section{Results \label{Results}}
Appendix \ref{StokesApp} reports the maps of the three Stokes parameters I, Q, and U. From these, we derive the polarised flux ($I_\mathrm{P}$), the polarised fraction ($P$) and the polarisation angle ($PA$), {following the standard equations:} 
\begin{align}
& I_\mathrm{P} = \sqrt{Q^2+U^2 -\sigma_{I_\mathrm{P}}^2} \; , \label{Derivation1}\\
& P =\frac{I_\mathrm{P}}{I} \; , \label{Derivation2}\\
& PA = \frac{1}{2} \arctan \left( \frac{U}{Q}\right) \; . \label{Derivation3} 
\end{align}
{In order to debias the polarised intensity, in Eq. \eqref{Derivation1} we have removed the contribution of the flux uncertainty ($\sigma_{I_\mathrm{P}}$). This procedure is necessary since $I_\mathrm{P}$ is biased to positive values, while the Stokes U and Q parameters can be positive or negative \citep{Vaillancourt06}. To avoid oversampling, after smoothing the data we re-gridded the maps to a pixel size of $9.1''$ ($\approx 4$ pixels per smoothed beam). Table \ref{ResultsValues} summarises the peak and mean values for the derived parameters.} Figure \ref{StokesIVec} shows the Stokes I emission (in colorscale), with polarization vectors already rotated by 90\degree (in black) to indicate the plane-of-sky component of the magnetic field. We mask data points with a signal-to-noise ratio in polarisation fraction lower than 3. 

\begin{table}[h]
\renewcommand{\arraystretch}{1.4}
\centering
\caption{  Peak and mean values of the polarised flux, polarised fraction and position angle. \label{ResultsValues}}

\begin{tabular}{cr@{$\pm$}lr@{$\pm$}l}
\hline
  Parameter                     & \multicolumn{2}{c}{   Peak}  &  \multicolumn{2}{c}{  Mean\tablefootmark{a}}  \\ 
\hline
$I_\mathrm{P}$ {(Jy/beam)}      & ${0.53}$&${0.07}$ & ${0.28}$&${0.07}$   \\
$P$ (\%)         & ${28}$&${8}$    & ${6.6}$&${1.8}$ \\
$PA$ (\degree) & \multicolumn{2}{c}{-}      & $45$&$7$    \\  \hline
\end{tabular}
\tablefoot{\tablefoottext{a}{  The mean values are computed over positions satisfying ${ \sigma_{P} / P    > 3.0}$ and ${P < 30 }$\%}}
\end{table}

\par
\section{Analysis and discussion}
\subsection{The magnetic field direction}
\begin{figure}
\centering
\includegraphics[width = 0.5\textwidth]{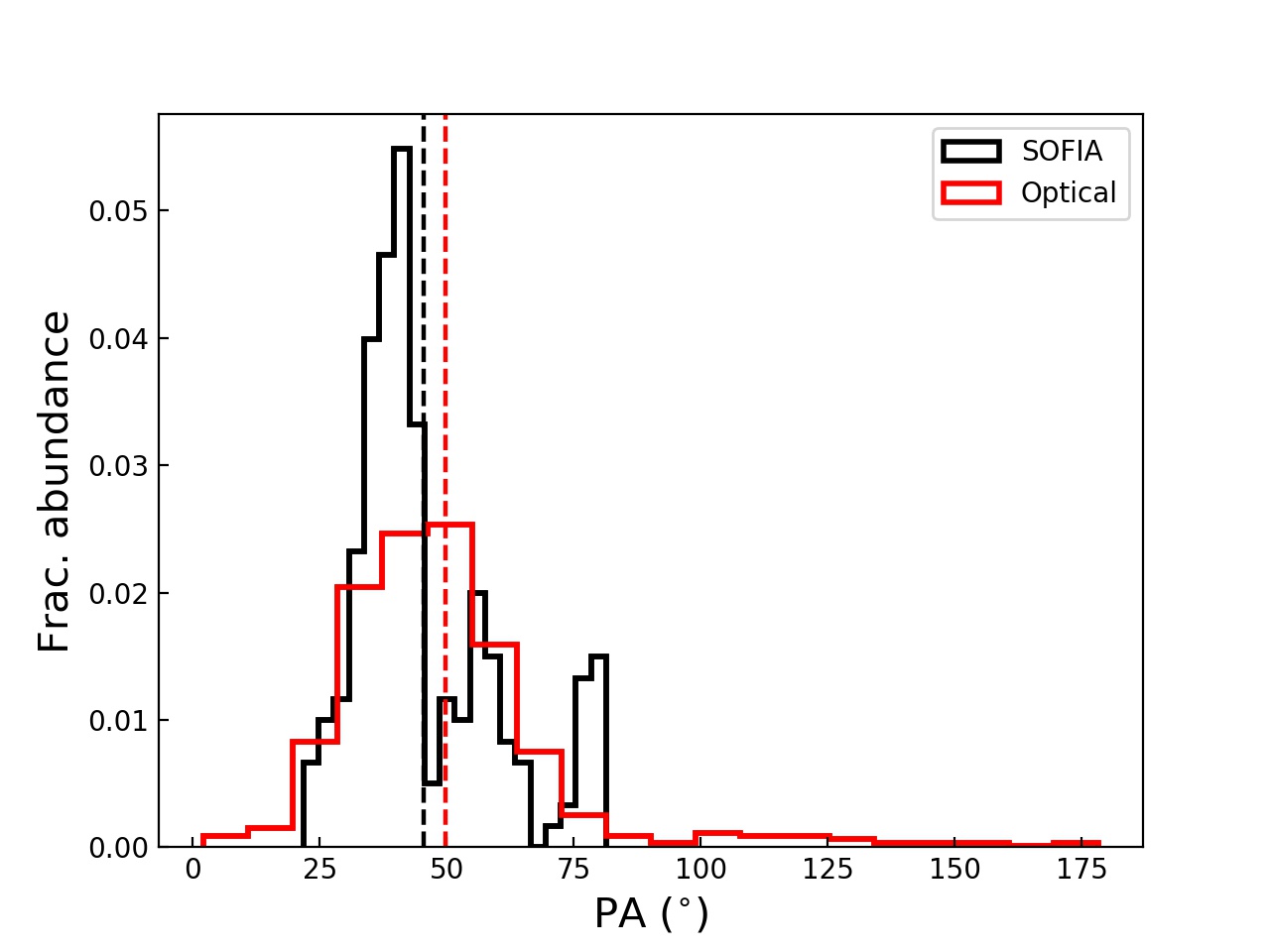}
\caption{Fractional abundance distribution of the magnetic field direction angles as traced by the THz data (black) and optical ones (red).The distributions mean values, shown with vertical dashed lines, are $45$\degree (SOFIA) and $50$\degree (optical). The SOFIA data present a three peaked distribution, with peaks being at $\approx 40$\degree, $\approx 55$\degree, and $\approx 77$\degree.  These correspond to vectors coming from the two halves of the {presumed} hourglass shape and the filament, respectively. \label{histo}}
\end{figure}

Figure \ref{StokesIVec} shows the magnetic field direction traced by the SOFIA data. The field at sub-parsec scale is aligned with the large (parsec) scale one observed in starlight optical polarisation. The histograms of the two position angle distributions are presented in Figure \ref{histo}. The mean values of the two distributions differ by $\approx 5$\degree, less than the mean uncertainty on the SOFIA position angles ($\langle \epsilon_{PA} \rangle = 7$\degree). Since the two datasets are sensitive to two different regimes, the optical data tracing large cloud scale, whilst the SOFIA ones being sensitive to the core scales, this means that the uniform magnetic field of the cloud has been inherited by the core, and that the gravitational collapse was magnetically-driven. \par
In Fig. \ref{StokesIVec} we also show that the outflow direction ($PA = 35\text{\degree}$, \citealt{Bjerkeli16}) lies almost parallel to the magnetic field, which has a mean direction of $\langle PA \rangle = (45\pm7)$\degree. This behaviour is consistent with the prediction of the theory of the magnetically driven collapse, according to which a strong magnetic field can efficiently remove the excess of angular momentum from rotating cores, thus aligning the rotation axis with the magnetic one (see e.g. \citealt{Li14}, and references therein). \par

Another prediction of the theory is the hourglass shape of the magnetic field lines. {In IRAS15398 this is not clearly visible, even though we do detect a bending of the field lines which may hint to a partial hourglass shape. In fact, on the side of the core facing South-East, the field lines pinch inwards toward the centre of the object.} The mean position angle North-East of the source is {$PA= (55\pm 7)$\degree}, whilst {$PA = (36\pm7)\text{\degree}$} is derived South-West of it. The change in angle is thus $19$\degree, significantly larger than the uncertainty. {We do not detect the North-West side of the hourglass}, which can be due to two main reasons. First of all, on that side we lack the sensitivity to detect \text{a bend in the field lines}, since the polarised flux is less prominent. This can be due to the presence of the filamentary structure that extends towards West. This filament provides more shielding from the external radiation field with respect to the other side of the core, which is more exposed. Since this radiation is responsible of the grain alignment, according to the RAT theory, grains are less efficiently aligned on the North-West side of IRAS15398, which results in a {lower} polarised flux. Moreover, the presence of the filament can affect the detected morphology of the field lines in a second way. In fact, the presence of the extended dust emission in the West direction can have played a role in disturbing the spherical collapse of the core. This scenario is supported by the few detections along the filament, which reveal a significant twist in the $B$ direction ($ PA  \approx 69$\degree). This suggests that in this portion of the source the magnetic field morphology has been perturbed, likely by accretion motions towards the central object.

\subsection{Depolarisation at high column densities}
\begin{figure}
\centering
\includegraphics[width = 0.5\textwidth]{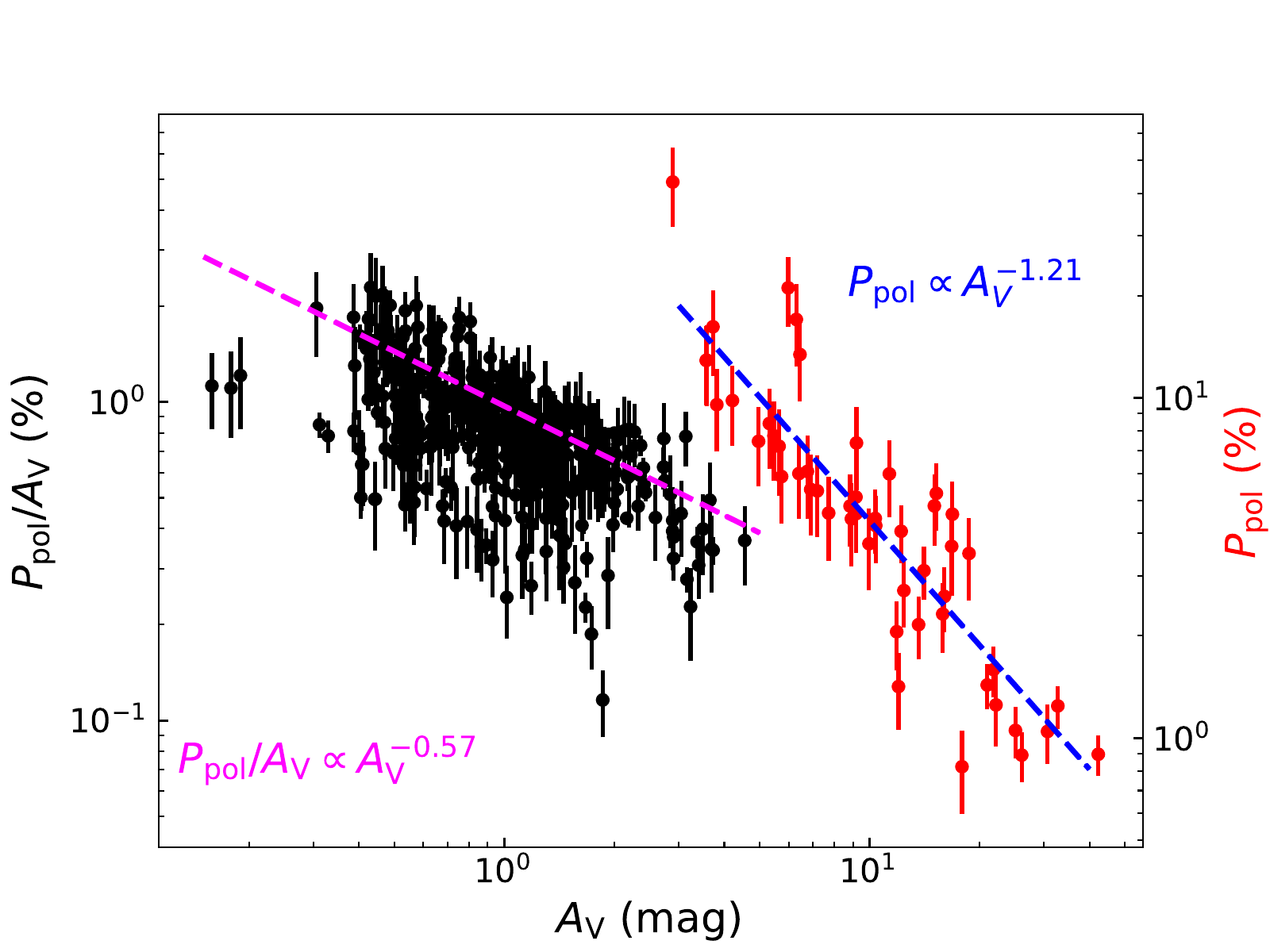}
\caption{Scatter-plots of the polarisation efficiency as a function of the visual extinction in magnitudes for the optical data (in black) and for the {FIR} ones (in red, {only one point every other is shown}), in logarithmic scale. The best fit to each data-set, as described in the main text, is shown with the pink and blue dashed curve, respectively. The best-fit equations are also shown in the bottom-left and top-right corners.
 \label{PpolAv}}
\end{figure}

Figure \ref{StokesIVec} shows that the emission is less polarised as the column density gets higher, towards the centre of the core. This effect has been widely observed in the literature \citep{Matthews09,Alves14, Jones15, Jones16}. In Figure \ref{PpolAv} we show the scatterplot of the polarisation fraction as a function of the visual extinction. {We also report the polarisation efficiency for the optical data, which is defined as the polarised fraction normalised by the visual extinction (see the Introduction in \citealt{Santos17}). In this way, one correctly takes into account the attenuation of the background stellar light. We use the data from \cite{Franco15}, limiting to the Galactic latitude range $16.2\text{\degree} < b < 17.0\text{\degree}$ (the "Middle" region in \citealt{Franco15}).} \par
There are a number of possible explanations for the depolarisation at high column densities. In general, if the alignment is dominated by radiative torque due to the interstellar radiation field, at higher visual extinction less and less radiation penetrates the core and dust grains are less aligned. Moreover, the alignment efficiency is highly sensitive to the dust grain population  \citep{Bethell07, Pelkonen09, Brauer16}. In particular, \cite{Brauer16} showed that polarization can decrease by up to 10\% when grains have sub-mm/mm sizes. Grain growth is expected in Class 0 objects, as shown by \cite{Jorgensen07, Kwon09, Chiang12}. Finally, geometrical smearing is possible. Close to the central object, {a highly perturbed magnetic field is likely to be present due to the dominant effect of gravity at the small scales which are unresolved by the SOFIA data.} 
Higher angular resolution observations are needed to disentangle these scenarios. \par
We fit a linear relation in the log-log space for the polarization efficiency vs visual extinction in each data-set, deriving the slope of the relation $P_\mathrm{eff} \propto A_\mathrm{V}^{- \alpha}$ (Figure \ref{PpolAv}). For the optical data, tracing the cloud scales, we find $\alpha = 0.57 \pm 0.07$, which is consistent with modelling of RAT alignment in molecular clouds (see e.g. \citealt{Whittet08}, who also report the measurements in Taurus and Ophiucus clouds). For the FIR data, we find a steeper slope ($\alpha = 1.21 \pm0.12$), implying a strong depolarisation for $10 \lesssim A_\mathrm{V} \, \text{(mag)}\lesssim 50$. Values of $\alpha \approx 1$ are often found in literature for prestellar and young Class 0 objects \citep{Alves14,Jones15,Jones16}. The different slopes exhibited by the optical and the {FIR} data suggest that {the two datasets are probing two different regimes in the grain alignment. This in turn can be related to two different grain populations}, with the larger grains in the core less efficiently aligned to the $B$-field than the ones in the cloud {(see e.g. Sect. 3 in \citealt{Andersson15})}, {or it can be due to a less efficient RAT, with the grains traced by SOFIA more shielded from the interstellar radiation field with respect to those traced by the optical data}.  Another possibility is a more entangled magnetic field lines, as mentioned before. According to recent results, a slope of $\alpha \approx 1$ can be due to biases introduced by the SNR cut \citep{Wang19}. To exclude this possibility, we fit the linear relation lowering the threshold to $\rm SNR > 0.5$, obtaining $\alpha = 1.26 \pm0.09$, in agreement the previous result, thus suggesting that no significant bias is introduced by the data selection.   
\subsection{{Angular dispersion function} and field strength \label{SFsec}}
{The angular dispersion function (ADF) method}, introduced by \cite{Hildebrand09} and \cite{Houde09}, involves the computation of the autocorrelation function of the position angles. {It allows to derive quantitative information on the B-field without depending on assumption on the field morphology. The method assumes that the magnetic field is composed by an ordered component ($B_0$), slowly changing spatially, and a turbulent or random component ($B_\mathrm{t}$), which is characterised by a coherent length scale $\delta$. We now consider the autocorrelation function of the position angles  $\Delta \Phi$, i.e. the difference in angle of every pair of vectors separated by the distance $l$:}
\begin{equation}
\langle \Delta \Phi^2 (l) \rangle = \frac{1}{N} \sum_{i=1}^N \left [   \Phi({x}) - \Phi ({x}+{l}) \right ]^2 \, .
\end{equation}
{\cite{Hildebrand09} showed that on scales $d$ such that $\delta < d << \Delta$, where $\Delta$ is the physical scale of the analysed source, the structure function has the form:}
\begin{equation}
\langle \Delta \Phi^2 (l) \rangle = b^2 +a^2 l^2 + \sigma_M^2 \, .
\label{SF_eq}
\end{equation}
{where $\sigma_M$ is the uncertainty on the measurements. The parameter $a$ is linked to the large-scale variations of the $B$-field, while $b$ is related to the ratio of the turbulent and uniform magnetic field components via $B_\mathrm{t}/B_0 = b/\sqrt{2-b^2}$.} \par
{\cite{Houde09} further expanded the analysis, in particular including the effects introduced by the telescope beam size. In case of a Gaussian beam with standard deviation $W$, they found the following expression:} 
\begin{equation}
\begin{split}
\langle \Delta \Phi^2 (l) \rangle = & 2 \sqrt{2 \pi} \left ( \frac{B_\mathrm{t}}{B_0}\right )^2 \frac{\delta^3}{(\delta^2 + 2 W^2) \Delta '} \times \\
& \times   \left [ 1- \exp \left (-\frac{l^2}{2 ( \delta^2 + 2 W^2)} \right )\right ] +m^2 l^2, 
\end{split}
\label{Houde_eq}
\end{equation}
{where $\Delta '$ is the cloud effective thickness, and $m$ is a parameter related to the large-scale structure of the magnetic field, which does not involve turbulence.  }\par
{In order to fit Eq. \eqref{Houde_eq} to derive the $B_\mathrm{t}/B_0$ parameter,  we compute $ \Delta \Phi^2 $ for all the available pairs of points with $0''<l <180''$, divided in 9 bins $20''$ wide. The choice of binning is determined by the angular resolution of the observations (we do not want to oversample the beam size) and by the map size (each bin should contain a fairly constant number of points). We compute the uncertainty $\sigma _\mathrm{M}$ in each bin propagating the uncertainties on the position angles, assumed to be uncorrelated. They are usually within 3-5\%. The full-with-half-maximum of SOFIA smoothed beam {(42$''$)} corresponds to {a standard deviation of} $W = 18'' = 13 \rm \, mpc$.}\par 
{Due to the limited number of data-points available for the fit, we fix the turbulence correlation scale on $\delta = 20 \rm \, mpc$. This value is in good agreement with previous estimation of this quantity in other star forming regions: for instance \cite{Houde09} found $\delta = 16 \rm \, mpc$ in OMC-1,  \cite{Frau14} derived $\delta = 13-33 \rm \, mpc$ in the high-mass star forming region NGC 7538, and \cite{Coude19} reported $\delta = 7 \rm \, mpc$ in Barnard 1. Furthermore, we assume the cloud thickness $\Delta '$ to be equal to the source effective radius ($r_\mathrm{eff}$), defined as the radius of a circular region of identical surface area. To compute the latter, we considered only positions with  $\rm Stokes\; I > 0.05 \,$Jy/pix. This threshold roughly corresponds to the lowest contour in $N\rm(H_2)$ in Figure \ref{StokesIVec} ($1\times 10^{22} \, \rm cm^{-2}$), and it is the first closed contour that comprises both the central core and the filamentary structure towards West. We obtain $r_\mathrm{eff} = 0.1 \rm \, pc$, a result which is also in agreement with the typical size of cores and filamentary structures in molecular clouds (see e.g. \citealt{Arzoumanian11, Andre14}). We explore later in the section the effects of these assumptions on the results.} \par
\begin{figure}
\centering
\includegraphics[width = 0.5\textwidth]{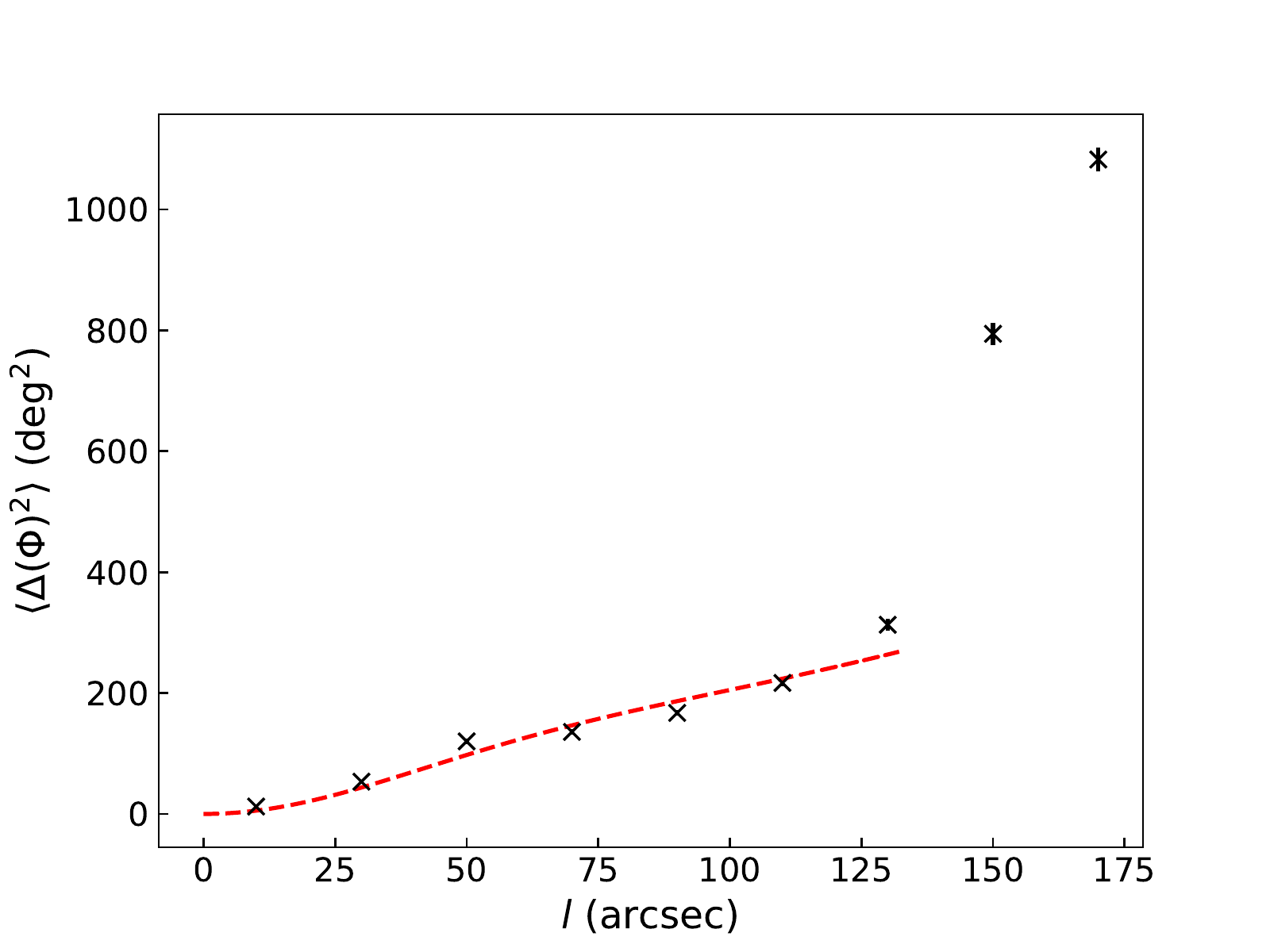}
\caption{{The angular dispersion function of the SOFIA data} with respect to the distance parameter $l$, computed as described in {Sect. \ref{SFsec}}. The measurement uncertainties $\sigma_\mathrm{M}$ are shown as error-bars, {and for low values of $l$ they are too small to be seen}. The best fit to the data points, following Eq. \eqref{Houde_eq}, is shown with a dashed curve. \label{SF}}
\end{figure}

{In Figure \ref{SF} we show the resulting data points and their uncertainties. We also show the best-fit solution found for Eq. \eqref{Houde_eq}. The obtained best-fit parameters are reported in Table \ref{FitSummary}. We emphasise that while figures and tables are expressed in deg/arcsec units, the actual fit is performed in rad/pc units.}\par
In order to estimate the strength of the magnetic field, we apply a modified version of the method proposed by \cite{ChandrasekharFermi53} (CF). The CF approach assumes the equipartition of kinetic and perturbed magnetic field energy to link the magnetic field strength on the plane of sky ($B_\mathrm{pos}$) to the velocity dispersion of the gas {($ {\sigma_\mathrm{V}}$)} and to the dispersion of the polarization angle ($\delta \phi$), which in turn corresponds to the turbulent ratio ($\delta \phi  \equiv B_\mathrm{t}/B_0$). Further development of the theory led to the equation:
\begin{equation}
B_\mathrm{pos} = \sqrt{{4} \pi \mu m_\mathrm{H} n_\mathrm{H_2}}\frac{{\sigma_\mathrm{V}}}{{\delta \phi}} \; ,
\label{Bfield}
\end{equation}
which is valid in the assumption of small polarization angle ($\delta \phi <25$\degree, \citealt{Ostriker01, Padoan01, Crutcher04}). In Eq. \eqref{Bfield}, $\mu$ is the gas mean molecular weight per hydrogen molecule, assumed to be $\mu = 2.8$ \citep{Kauffmann08}, $m_\mathrm{H}$ the hydrogen mass ($m_\mathrm{H} = 1.67\, 10^{-24}$g), {and} $n_\mathrm{H_2}$ is the gas volume density. \par
To compute $B_\mathrm{pos}$, we need to determine the quantities $n_\mathrm{H_2}$ and ${\sigma_\mathrm{V}}$. For the gas volume density, we consider the \hh column density map, {integrated over the region used to derive the effective radius}, obtaining the total number of \hh molecules $H_\mathrm{tot}$ in the considered region. Assuming a uniform distribution of the gas, we can thus derive:
\begin{equation}
n_\mathrm{H_2} = \frac{3}{4 \pi} \frac{H_\mathrm{tot}}{r_\mathrm{eff}^3} = 2.6\times 10^4 \, \mathrm{cm^{-3}}\; .
\end{equation}
For the gas velocity dispersion, we use the results of \cite{Benedettini12}, who observed several molecular tracers in Lupus I with the MOPRA telescope, which at $3\,$mm has a beam size similar to the one of our data. At the position of IRAS15398, they report {full width at half maximum ($FWHM$) values} in the range $0.31-0.72\, \mathrm{km\, s^{-1}}$. We adopt the mean of their results {(excluding the CS (2-1) line, which is most likely optically thick)}, $FWHM = 0.41\,\mathrm{km\, s^{-1}}$, {corresponding to a velocity dispersion $\sigma_\mathrm{V} = 0.17\,\mathrm{km\, s^{-1}}$}. Inserting these values in Eq. \eqref{Bfield}, we obtain $B_\mathrm{pos} = {78}\,\mu$G. Due to the strong assumptions of the method, such as the core uniform density, we expect this value to be accurate within a factor of two. Furthermore, we want to highlight that it is intrinsically a lower limit, since it neglects the line-of-sight component of the field.  \par
A fundamental parameter is the mass-to-flux ratio ($M/\Phi$), which is determined by the ratio between the gravitational and the magnetic energy. It provides the dynamical state of the core (equilibrium or collapse). In particular, it is interesting to compute its observed value with respect to the critical value, defined by the parameter $\lambda$ \citep{Crutcher04}:
\begin{equation}
\label{lambda}
\lambda = (M/\Phi)_\mathrm{obs} /(M/\Phi)_\mathrm{crit} = 7.6 \times 10^{-21} \left( \frac{N\mathrm{(H_2)}}{\mathrm{cm^{-2}}}  \right) \left ( \frac{B}{\mu \mathrm{G}} \right )^{-1} \; ,
\end{equation}
where $N(\rm H_2)$ is the mean column density of the source, computed in the region over which B is measured. Focusing on  the same region used to derive $n_\mathrm{H_2}$, we obtain $N \mathrm{(H_2)} = 9.8 \times 10^{21}\,\rm cm^{-2}$, and therefore $\lambda ={0.95}$. {Our derived value of $\lambda$ suggests that the core is in transitional state between being subcritical and supercritical}. \par
\begin{table}[h]
\renewcommand{\arraystretch}{1.4}
\centering
\caption{Summary of the main parameters concerning the magnetic field, derived as described in Sec. \ref{SFsec}. \label{FitSummary}}
\begin{tabular}{cccc}
\hline
 $B_\mathrm{t}/B_0 $ &  $m$ ($\rm deg^2 / arcsec^2$) & $B_\mathrm{pos}$ ($\mu$G) & $\lambda$\\
\hline
$0.267 \pm 0.007$  & $170 \pm 50$ & ${78}$& ${0.95}$  \\
\hline
\end{tabular}
\end{table}

{Our analysis of the mass-to-flux ratio considers only the gas mass of the source, and neglects the protostellar mass. However, the total mass within the contour $\rm Stokes\; I > 0.05 \,$Jy/pix is $\approx 6 \, M_\odot$, significantly larger than the the protostellar mass (the upper limit is $M_* < 0.1 M_\odot$, \citealt{Yen17}). This approximation therefore causes an error of less than $2$\%, well below our uncertainties.}

\subsection{  Influence of the assumed parameters}
{The analysis described above depends on the assumption of several key parameters. We explore in this section how the results are affected by changing the assumed values. The first assumption regards the turbulence correlation scale, $\delta$. The literature results seem to suggest that $\delta$ is lower for low-mass star-forming regions, with respect to high-mass ones: both \cite{Liu19} and \cite{Coude19} found $\delta  < 10 \, \rm mpc$ in low-mass cores. We therefore tested other 3 values ($5, 10, 15 \rm \, mpc$), obtaining $B_\mathrm{pos} = {16, 40, 62} \, \rm \mu G$ respectively. The magnetic field strength hence increases with increasing $\delta$. The turbulent-to-uniform component ratio is lower than one in all cases but the extreme one ($\delta = 5\rm \, mpc$), for which is marginally higher than one ($B_\mathrm{t} / B_0  = 1.3$). {On the other hand, the mass-to-flux ratio for these values is $\lambda > 1.0$, pushing the core's state more towards the supercritical regime.}  \par
{A second assumption that is made regards the cloud effective thickness ($\Delta '$), which may not necessary correspond to the real thickness $\Delta$. In particular, according to \cite{Houde09}, $\Delta ' \le \Delta$, where the equality holds only if the large-scale magnetic field is completely uniform. We therefore repeated the analysis using $\Delta ' = r_\mathrm{eff} /2$. The results vary of less than $30$\%, below our uncertainties.  } \par
{The magnetic field strength can also be calculated directly with the modified Chandrasekhar-Fermi method (Eq. \eqref{Bfield}), computing directly the standard deviation of the position angles ($\delta \phi$){, hence avoiding} the assumptions made in the ADF method. {In this case, Eq. \eqref{Bfield} must be corrected for a numerical factor ($Q$), in order to take into account that its original version tends to overestimate the magnetic field \citep{Ostriker01, Crutcher04}. A usually adopted value is $Q=0.5$.} From the SOFIA data, we measure $\delta \phi _\mathrm{obs} = 14$\degree. This value is however biased upward by the observational uncertainties on the position angles ($\epsilon_{PA}$), and it must be corrected using $\delta \phi^2 =  \delta \phi _\mathrm{obs}^2 - \epsilon _{PA} ^2 = (12\text{\degree})^2$ \citep{Crutcher04}. Hence, introducing $\delta \phi = 12$\degree in Eq. \eqref{Bfield}, we obtain $B_\mathrm{pos} = {47} \, \rm \mu G$, which is in well agreement with the result of the full ADF analysis, given the uncertainties. We conclude that the results are robust with respect to our assumptions. }

\section{Conclusions}
We have performed polarimetric observations of the dust thermal emission at $1.4\,$THz in the class 0 protostar IRAS15398. Our data unveil a very ordered magnetic field. The direction of the magnetic field is consistent within uncertainties with the one found by \cite{Franco15} using optical data, which trace large cloud scales. This suggests that during its evolution the core preserved the $B$-field morphology inherited from the parental cloud, and experienced a magnetically driven collapse. \par
The magnetic field lines on the South-East side of the core pinch inwards, {hinting to} the hourglass shape predicted by the theoretical models. {However, we lack the sensitivity to clearly identify the hourglass shape.} The lack of such a pinching on the other side of the core may be linked to the presence of an extended structure linked to the central core. The polarisation vectors detected in this filament show a significantly different direction of the magnetic field, perhaps hinting that material is being dragged towards the central object and that the infalling gas is perturbing the magnetic field morphology. Further line observations, which allow to infer the velocity field in the source, will help us enlighten this point. \par
Using the modified CF method we estimate the magnetic field strength on the plane of sky {to be  $B_\mathrm{pos} = {78}\,\mu$G}, accurate within a factor of two. This value is  {in the lower end of the range found} on similar spatial scales performed in star forming regions for example with the JCMT telescope (see e.g. \citealt{Crutcher04,Kwon18,Soam18,Liu19}), where $B$-strengths of $80-5000\,\mu$G have been reported. \par
{The mass-to-flux ratio that we derive ($\lambda = {0.95}$) is {close to the transcritical regime ($\lambda = 1$).} Usually cores are found to be super-critical (see e.g. \citealt{Troland08}), in contrast with molecular clouds which are often subcritical: for instance, \cite{Franco15} found $\lambda = 0.027-0.057$ in Lupus I. {IRAS15398 hosts a protostellar object, which indicates that indeed gravitational collapse has happened and that, on some scales, the source must be supercritical. Our observations are likely missing the angular resolution to probe those scales, and therefore we are seeing an intermediate state due to the subscritical surrounding medium.} {On the other hand,} the uniform-to-turbulent ratio is smaller than one  ($B_\mathrm{t}/B_0  = 0.267$), suggesting a strongly magnetised core.} This scenario is also supported by the lack of a large Keplerian disk in the source (\citealt{Yen17} found an upper limit to the disk size of $30\,$AU). In fact, magnetic braking is an efficient way to remove angular momentum from the infalling and rotating material and prevent the formation of large disks \citep{Li14}. \par
Overall, our data suggest that IRAS15398 evolved in an highly magnetised environment, and that the ordered magnetic field was preserved from cloud scales down to core scales. Future polarimetric observations {with ALMA} at high resolution will allow us to investigate wether this uniform magnetic field is preserved down to the disk/envelope scales.

\begin{acknowledgements}
Based on observations made with the NASA/DLR Stratospheric Observatory for Infrared Astronomy (SOFIA). SOFIA is jointly operated by the Universities Space Research Association, Inc. (USRA), under NASA contract NNA17BF53C, and the Deutsches SOFIA Institut (DSI) under DLR contract 50 OK 0901 to the University of Stuttgart. This research has made use of data from the {\em Herschel} Gould Belt survey (HGBS) project (http://gouldbelt-herschel.cea.fr). The HGBS is a {\em Herschel} Key Programme jointly carried out by SPIRE Specialist Astronomy Group 3 (SAG 3), scientists of several institutes in the PACS Consortium (CEA Saclay, INAF-IFSI Rome and INAF-Arcetri, KU Leuven, MPIA Heidelberg), and scientists of the {\em Herschel} Science Center (HSC). {The authors thank the anonymous referee for his/her helpful comments.}
\end{acknowledgements}

\appendix

\section{The Stokes parameters \label{StokesApp}}
Figure \ref{Stokes} reports in the three panels the Stokes parameters observed with SOFIA. The maps have been smoothed to $42''$ with respect to a native resolution of $19''$ to improve the SNR.
\begin{figure}[h]
\centering
\includegraphics[width = 0.5\textwidth]{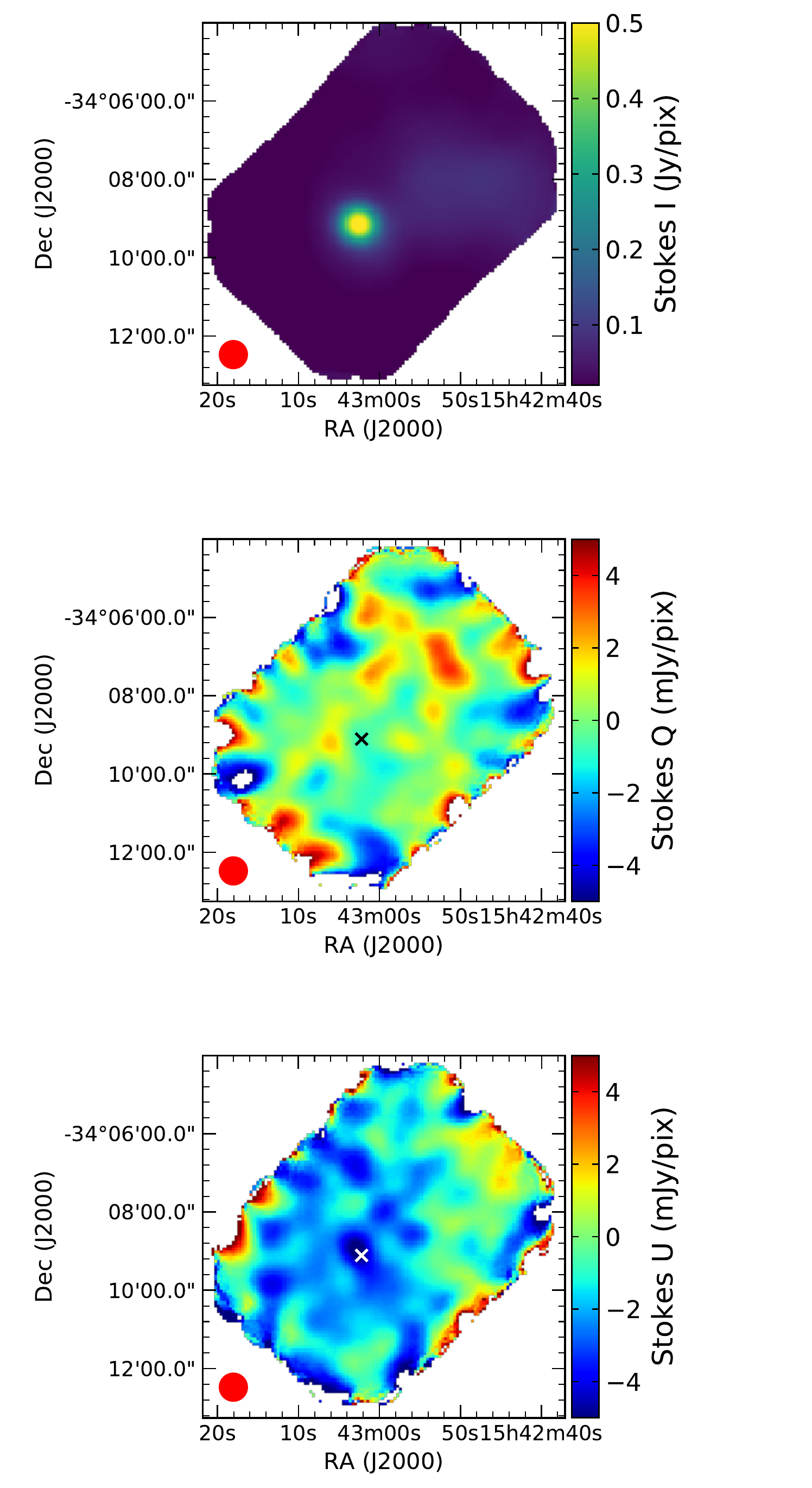}
\caption{Colormap of the three Stokes parameter I, Q, and U (from top to bottom). The smoothed beam size is indicated in the bottom-right corners. {In the central and bottom panels, the cross represents the position of the protostar.} \label{Stokes}}
\end{figure}

\end{document}